\documentclass[a4paper]{article}

\usepackage{INTERSPEECH2021}
\usepackage{multirow}
\title{Deep Representation Decomposition for Rate-Invariant Speaker Verification}
\name{Fuchuan Tong$^{\dag 1}$, Siqi Zheng$^3$, Haodong Zhou$^{1}$, Xingjia Xie$^1$, Qingyang Hong$^{2}$, Lin Li$^{1}$
\thanks{\dag  The work was done during an internship at Alibaba.}
\thanks{This research is funded by the National Natural Science Foundation of China (Grant
No. 61876160 and No. 62001405) and Fundamental Research Funds for the Central Universities (No. 20720210087).}
}
\address{
  $^1$ School of Electronic Science and Engineering, Xiamen University, China \\
  $^2$ School of Informatics, Xiamen University, China \\
  $^3$ Speech Lab, Alibaba Group
  }

\email{\{qyhong, lilin\}@xmu.edu.cn}
\begin{document}

\maketitle

\begin{abstract}
  While promising performance for speaker verification  has been achieved  by deep speaker embeddings, the advantage  would reduce in the case of speaking-style variability. Speaking rate  mismatch is often observed in practical speaker verification systems, which may actually degrade the system performance. To reduce intra-class discrepancy caused by speaking rate,  we propose a deep representation decomposition approach with adversarial learning to learn  speaking rate-invariant speaker embeddings. Specifically, adopting an attention block, we decompose the original embedding into identity-related component and rate-related component through multi-task training.  Additionally, to  reduce the latent relationship between the two decomposed components, we further propose a cosine mapping block to train the parameters adversarially to minimize the cosine similarity between the two decomposed components. As a result, identity-related features become robust to speaking rate and  then are used for verification. 
   Experiments are conducted on VoxCeleb1  data and  HI-MIA data to demonstrate the effectiveness of our proposed approach.
  
\end{abstract}

\section{Introduction}

Speaker verification is  a typical biometric authentication technology that verifies the identities of speakers from their voices.
Recently,  deep learning based speaker verification algorithms have achieved excellent performance. 
The main task of deep  embedding based speaker recognition system involves extracting a high dimensional embedding from an utterance that characterizes each speaker uniquely. Recently, deep speaker embeddings (x-vectors) extracted from time delay neural network (TDNN) \cite{snyder2018x} or other encoders (e.\,g., Res2Net \cite{zhou2020resnext}, ECAPA-TDNN \cite{desplanques2020ecapa}) have become state-of-the-art for speaker verification.  In terms of `in-the-wild' environments and speaking-style variability, however, the embeddings will include speaker-unrelated information, 
which will significantly degrade their performance. Therefore, improving  the robustness of speaker embeddings has  become a crucial research issue for the `in-the-wild' speaker verification task.

Most recently, a few works were proposed to  eliminate the impact of identity-unrelated information. For example, Peri \textit{et\ al.} \cite{peri2020robust}
adopted an adversarial invariance architecture to train a network and  extracted robust speaker-discriminative representations.
Tai \textit{et\ al.} \cite{tai2020seef} 
proposed the SEEF-ALDR  framework to minimize the impact of identity-unrelated information via adversarial learning based disentangled representation. Similarly, Kwon \textit{et\ al.} \cite{Kwon2020IntraclassVR} improved the  SEEF-ALDR  framework and disentangled identity-related and identity-unrelated information using a mutual information criterion through an auto-encoder framework.

In this work, we focus on the speaking rate mismatch problem. This problem occurs when a speaker enrolled in a normal speaking rate but tested with a slower or faster speech utterance. Speaking rate mismatch is often observed in practical speaker verification systems.  For example, people tend to speak faster when they are in a hurry, while speaking slowly due to exhaustion or illness. 
Small differences in speaking rate between enrollment and test utterances will not be a problem, but severe mismatch will lead to serious performance degradation.
Zeng et al. \cite{zeng2015learning} found that some speaker information will be discarded in fast speech, while slow speaking rate damages the spectrum of speech signals. 
In early works, Grimaldi \textit{et\ al.} \cite{ grimaldi2009speech} studied the  impact of speaking rate on speaker verification systems and confirmed that verification performance degraded due to speaking rate mismatch. Heerden \textit{et\ al.} \cite{van2007speech} used the phoneme duration as an additional feature and augmented it to the conventional Mel-frequency cepstral coefficients (MFCCs)  to mitigate the impact of speaking rate mismatch. Rozi \textit{et\ al.} \cite{rozi2016feature} proposed a feature transform approach that projected the speech features with slow speaking rate to those with normal speaking rate. 

In this paper, we propose a method to effectively decompose the original embedding into two uncorrelated components: identity-related component and rate-related component.  We utilize a rate estimation task  with a channel-wise attention block to obtain the rate-related feature and then disentangle it from the whole original embedding. Besides, to further reduce the latent relationship between the two decomposed components, we adopt a cosine similarity loss that minimizes  the cosine similarity between the identity- and rate- related features  in an adversarial manner. Specifically, a cosine mapping block is introduced to find the maximum similarity between the two components, while the encoder network and attention block aim to  reduce the similarity adversarially. Through the feature decomposition  and adversarial training,  the rate information can be significantly removed.
Our proposed  framework can be  implemented as a simple yet effective residual branch by integrating it into existing encoders while only adding a tiny amount of parameters and computation.  


\section{Related works}
\subsection{Time-scale modification}

Given an audio signal $x(t)$, we can alter its rate  to $x(\alpha t)$ by re-sampling and time-warping with a scale $\alpha$. However, this simple method will change the audio fundamental frequency and result in a different pitch.  This can be seen from the frequency domain, supposing the Fourier transform of $x(t)$ is $X(w)$, then the Fourier transform of $x(\alpha t)$ will be $\alpha^{-1}X(\alpha^{-1}w)$, which means the time-warping scale produces frequency components shifting.  These changes result in the audio sound as if they are uttered by different people. Lee \textit{et\ al.} \cite{lee2020nec} adopted this method for data augmentation and assigned new speaker labels to the augmented examples. That is to say, this method cannot be used to simulate different speech rates for the same person, which requires the same pitch content.

Thanks to the time-scale modification (TSM) technology, we can apply it to simulate diverse speaking rate scenarios. The TSM  alters the duration of an audio signal while retaining its local frequency content without affecting the pitch content and the prosody information \cite{ma2018short}. That is to say, the duration of the original audio  can be increased or decreased, but the important speaker identification features  remain unchanged. By doing so, the TSM  makes the modified audio sound  as if the speaker is talking at a slower or faster pace. Current methods for TSM can be roughly categorized into three categories: frequency domain, time domain, and hybrid methods.  In general, time-domain methods
are more effective at scaling transient signals, while frequency-domain methods excel in scaling harmonically complex audio. Hybrid methods leverage the strengths of frequency and time domain methods to produce higher quality results \cite{2006.00848}. Recently, in the field of  speech recognition and language identification, several works \cite{ma2018short,ko2015audio}  have adopted the TSM based algorithm for data augmentation.

\subsection{Latent identity analysis}


The latent identity analysis model can infer the latent variable through a statistical model from  the given observations. The general  formulation for latent identity analysis can be formulated as:
\begin{equation}
  \varPhi =\mu +\sum_{i=1}^n{\mathbf{U}_i\mathbf{x}_i}
  \label{eq3}
\end{equation}
where $\varPhi \in \mathbb{R} ^{d\times 1}$ refers to the speaker representation, $\mu$ is the mean of all the representations, the columns of $\mathbf{U}_i\in\mathbb{R}^{d\times m_i}$ span the subspace of different variation and $\mathbf{x}_i\sim N\left(\mathbf{0},\mathbf{I}\right)$ denotes the latent variable. 

 Recently, a few works \cite{villalba2017tied,kang2019adversarially} have applied  the latent variable model to speaker verification based on the variational autoencoder.
However, how to smoothly apply latent variable model to extract robust speaker embeddings still remains to be explored.


\subsection{Adversarial learning}
Adversarial training technique has greatly facilitated speaker verification in domain mismatch situations.
In speaker verification, the domain adversarial network usually consists of an encoder, a speaker discriminator, and a domain discriminator.
The encoder aims to fool the domain discriminator through learning source embeddings  that  resemble the target embeddings, while the domain discriminator aims to discriminate the learned embeddings from target domain. 
By this minimax game between the encoder and domain discriminator, the encoder can successfully minimize the distance between the source and target data distribution in the feature space.  
Besides, adversarial learning has been widely explored to improve the speaker embedding robustness for noisy \cite{peri2020robust,meng2019adversarial,zhou2019training}, cross-channel \cite{wang2018unsupervised,chen2020channel,luu2020channel}, and short utterances \cite{liu2020text} situations.

Unlike the general adversarial learning framework, we adopt a simpler adversarial training approach, i.e., minimizing the cosine distance between two features to reduce their correlation.


\section{Proposed methods}

\begin{figure}[t]
  \centering
  \includegraphics[width=\linewidth]{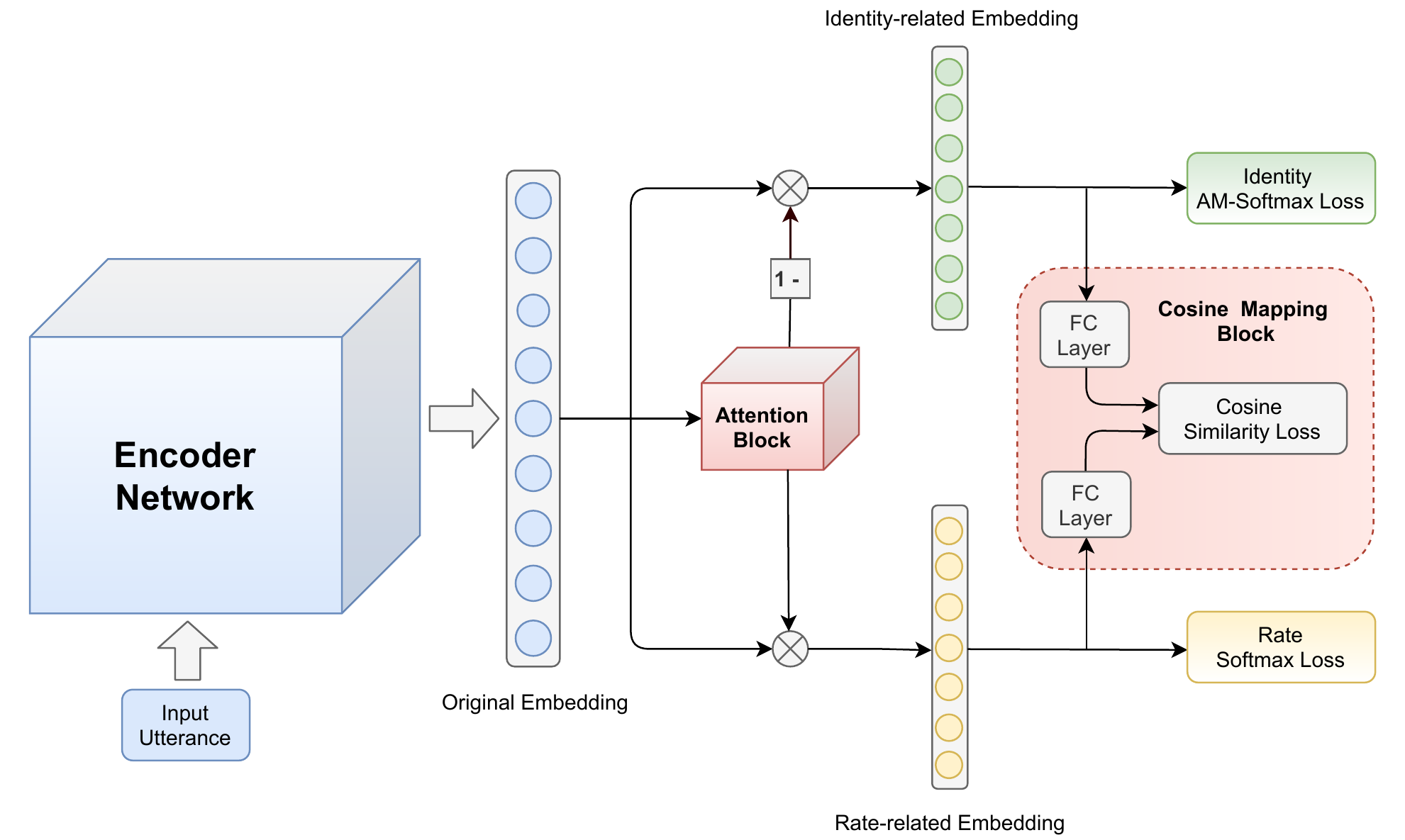}
  \caption{An overview of the proposed method.}
  \label{fig:frequency}
\end{figure}

\subsection{Feature decomposition}
As audio contains intrinsic identity information and other information, they can be jointly represented by the identity-related feature and variability-related feature. Thus,  we can decompose the two features from the original embedding $\varPhi$ in a supervised manner:

\begin{equation}
\mathbf{x}_{id}=\mathbf{\tilde{V}}\varPhi \,\,, \,\,\mathbf{x}_{rate}=\mathbf{\tilde{U}}\varPhi 
  \label{eq4} 
  \end{equation}
where, $\mathbf{x}_{id}$ is the identity-related feature, and $\mathbf{x}_{rate}$ is the rate-related feature.
$\mathbf{\tilde{V}}$ and $\mathbf{\tilde{U}}$ are the projection matrices, respectively. However, the identity-related feature is latent variable, implementing $\mathbf{\tilde{V}}$ and $\mathbf{\tilde{U}}$ in the network directly will be complicated. Since the identity-related feature is what we need in the rate-invariant speaker verification problem, and the rate-related feature can be easily obtained through a rate estimation task, we utilize a rate estimation task combining  with a channel-wise attention block \cite{hu2018squeeze,zhou2019deep} to disentangle the rate-related features from the embedding. More precisely, every input utterance is labeled by speaker-id and rate-id ($slow$, $normal$, and $fast$), and the identity estimation and rate estimation tasks update the parameters of the network simultaneously. Besides, an attention model is applied to perform dynamic channel-wise feature recalibration for the multi-task learning. Figure 2 shows an overview of our proposed method. 
We base on the  assumption that if the  reinforced features learned by the attention block are helpful for the variability estimation task, they should be suppressed from the identity-related features. Thus, the decomposition of the original embedding can be defined as:
\begin{equation}
\mathbf{x}_{id}=\left( 1-\mathrm{\sigma}\left( \varPhi \right) \right) \odot \varPhi \,\,, \,\, \mathbf{x}_{rate}=\mathrm{\sigma}\left( \varPhi \right) \odot \varPhi 
\label{eq5}
\end{equation}
where $\mathrm{\sigma}$  represents an attention weight learned by the attention model, and $\odot$ represents element-wise multiplication.



\subsection{Cosine similarity adversarial learning}
 The disentangled features, i.\,e., $\mathbf{x}_{id}$ and $\mathbf{x}_{rate}$, obtained through the attention mechanism, however, may  have some latent relationship with each other. That is to say,  $\mathbf{x}_{id}$ may still contain the speaking  rate related information which degrades the performance of speaker verification. To better guide the  rate information  disengaged from  the original embedding, inspired by \cite{wang2019decorrelated},
we further consider minimizing the cosine similarity between $\mathbf{x}_{id}$ and $\mathbf{x}_{rate}$.  Intuitively,  if the features of the two subspaces have low cosine similarity, their intrinsic latent relationship would also be very small. As shown in Figure 2,  the cosine similarity loss is computed between the two decomposed features by the cosine mapping block. Precisely, we evaluated the cosine similarity loss as follow:
\begin{equation}
  L_{cos}=\left(\overline{\mathbf{x}'}_{id}\cdot \overline{\mathbf{x}'}_{rate} \right) ^2
  \label{eq6}
  \end{equation}
where $\overline{\mathbf{x}'}$ is the normalized version of the fully-connected (FC) layer output  embedding $\mathbf{x}'$, 
 and  $\cdot$ represents dot production.  The square function is used to constrain $L_{cos}\in\left[ 0,1 \right]$.
Adopting adversarial learning, we divide the training process into two phases, termed as the cosine similarity maximization and minimization, respectively. That is to say, we first maximize the cosine similarity by training the cosine mapping module while freezing the encoder and attention module. Then, with the cosine mapping module fixed, cosine similarity is minimized along with the identity and rate estimation tasks by updating the encoder and attention module. By doing so, it plays a minimax game  during the adversarial training procedure. Finally, it renders the two features almost  orthogonal and further reduces the correlation between them.

Overall, the total objective function for the multi-task is formulated as:
\begin{equation}
  L=L_{id}+\lambda _1L_{rate}+\lambda _2L_{cos}
  \label{eq7}
  \end{equation}
where $L_{id}$ is the additive margin (AM) softmax  loss for the identity estimation task, $L_{rate}$ denotes the rate estimation softmax loss, $\lambda_1$ and $\lambda_2$ are scalar hyper-parameters to balance  these three losses.

\section{Experiments}
\subsection{Dataset}
To demonstrate the effectiveness of our proposed method, a dataset with different speaking rates is required and the amount of samples should be large enough to avoid overfitting. However, the size of available datasets to meet the experimental requirements is limited. In order to assess our systems’ robustness to speaking rate, we first created a large simulated dataset with different speaking rates based on the VoxCeleb1 dataset \cite{nagrani2020voxceleb}. Then, experiments were conducted on both  the augmented VoxCeleb1 simulated dataset and the HI-MIA \cite{qin2020hi} real corpus. 

The training set of VoxCeleb1  includes 148,642 recordings uttered by 1,211 speakers, while the test set consists of 37,720 test trials, including 4,878 utterances from 40 speakers. The utterances in the VoxCeleb1 dataset were collected from online video,  although not all of them were pronounced at a strictly normal speed, we assumed they were  normal speaking rate utterances and labeled them as \emph{normal}. To simulate different speaking rates from the original data,  we adopted the FFmpeg  libraries \cite{tomar2006converting} based TSM  algorithm  to modify the original audio ${x(t)}$ to ${x(\alpha t)}$ with a scaling speed factor $\alpha$. Compared with  the original speech, if $\alpha<1$, a slower speech was generated and labeled as \emph{slow}, if $\alpha>1$, a faster speech was generated and labeled as \emph{fast}. In our experiments, the scale $\alpha$ ranged from 0.5 to 2.0 and the change step was set to 0.1. For the training set, we added one-fourth of the original data for each scale $\alpha$ less than 1.0 (0.5--0.9), while for each scale $\alpha$ which is greater than 1.0 (1.1--2.0), one-eighth of the original data was added.  Thus, after being augmented by different speaking rates, the total training data size was about 3.5 times of the original data. Meanwhile, except for the original test trials, another 15 sets of test trials were also created, which were enrolled in normal speaking rate while tested in different specific speaking rates w.\,r.\,t a given $\alpha$. 


\begin{table*}[htbp]
  \centering
  \caption{Comparisons on VoxCeleb1  under different speaking rate test sets, the best results are bolded (EER[\%]).}
  \resizebox{\linewidth}{!}{
    \begin{tabular}{l|cccccccccccccccc}
      \hline
      \multirow{2}{*}{Systems} & \multicolumn{16}{c}{Speaking rate  scale $\alpha$} \\
\cline{2-17}   & 0.5   & 0.6   & 0.7   & 0.8   & 0.9   & 1.0   & 1.1   & 1.2   & 1.3   & 1.4   & 1.5   & 1.6   & 1.7   & 1.8   & 1.9   & 2.0  \\
    
      \hline
      \hline
      S1:Baseline & 5.84  & 5.15  & 4.75  & 4.32  & 4.12  & 3.99  & 4.01  & 4.05  & 4.27  & 4.41  & 4.69  & 5.02  & 5.14  & 5.56  & 5.86  & 6.21  \\
      S2:TSM aug.   & 3.98  & 3.82  & 3.73  & 3.69  & 3.68  & 3.58  & 3.61  & 3.69  & 3.72  & 3.73  & 3.89  & 3.99  & 4.01  & 4.17  & 4.33  & 4.45  \\
      S3:FD att.\  & 3.34  & 3.22  & 3.21  & 3.18  & 3.18  & 3.14  & 3.16  & 3.17  & 3.28  & 3.36  & 3.39  & 3.52  & 3.55  & 3.75  & 3.76  & 3.83  \\
      S4:AL cos. & 3.35  & 3.28  & 3.26  & 3.14  & 3.11  & 3.13  & 3.14  & 3.14  & 3.26  & 3.20  & 3.26  & 3.32  & 3.48  & 3.62  & 3.78  & 3.83  \\
      S5:FD-AL & \textbf{3.13} & \textbf{3.13} & \textbf{2.99} & \textbf{2.92} & \textbf{2.90} & \textbf{2.92} & \textbf{2.90} & \textbf{2.94} & \textbf{3.02} & \textbf{3.07} & \textbf{3.07} & \textbf{3.13} & \textbf{3.33} & \textbf{3.47} & \textbf{3.75} & \textbf{3.78} \\
      \hline

    \end{tabular}%
  \label{tab:addlabel1}%
  }
\end{table*}%

For the real scene speaking rate, experiments were conducted on the text-dependent speaker recognition database, HI-MIA. The phrase of HI-MIA is the wake-up words `Hi, Mia' in Chinese. The data was collected in a real home environment using microphone arrays and a high-fidelity microphone. The recordings of each speaker could be categorized into three subsets according to the speaking speed (i.\,e., slow, normal, and fast). 
Considering that we focused on examining the effect of different speaking rates, other factors (e.\,g., far-field, cross-channel) should be excluded. Thus, we only selected close-talking microphone utterances for the creation of the disjoint training set (254 speakers) and test set (42 speakers). Then, three  sets  of trials with different speaking rates were created as the cross-product of the utterances in the test set. In other words, all possible target and impostor samples were created from the test set.


\subsection{Experimental setting}

For all of the audios,  we extracted 40-dimensional MFCCs with a 25ms window and a 10ms frame shift, and an energy-based voice active detection (VAD) was conducted to filter out non-speech frames. Then, cepstral mean normalization (CMN) over a 3s sliding window was applied. 

For the  network  implementation, we applied the extended time delay network (E-TDNN) describe in \cite{snyder2019speaker} as the encoder network. 
Our proposed framework was applied by integrating it into the segment-level. We utilized the cross-entropy loss for speaking rate prediction and the AM-Softmax loss for speakers classification. 
We alternately run the cosine similarity maximizing process for 20 iterations and then switched to minimizing process for 50 iterations.
The empirically setting of hyper-parameters $\lambda_1$ and $\lambda_2$ in Equation 5 were: $\lambda_1=0.1$, $\lambda_2=0.1$.

Since we  had only limited real-world data with different speaking rates, the model trained directly on these data would most likely to be overfitting. Therefore,  we resorted to the transfer learning strategy for HI-MIA evaluations. Specifically, we utilized the models pre-trained on the VoxCeleb1 dataset for parameters initialization. 

At testing stage,  we computed the probabilistic linear discriminant analysis (PLDA) scoring and used the EER as the performance metric. All of these systems were implemented in ASV-Subtools \cite{tong2021asv}. 


\subsection{Evaluations on VoxCeleb1}
\label{sub:vox}
In this section, experiments were conducted on the Voxceleb1 dataset. Five different systems were designed for comparisons, listed as below: 
\begin{itemize}
\item S1: The baseline E-TDNN trained on the VoxCeleb1 training data.
\item S2: The network trained on TSM simulated and original  data (TSM aug.). 
\item S3: The feature decomposition system  with attention block (FD att.).
\item S4: The network  trained with cosine similar adversarial learning (AL cos.).
\item S5: The system combined with feature decomposition and cosine similar adversarial learning (FD-AL).
\end{itemize}

Comparisons between these systems under different speaking rate test sets are shown in Table~\ref{tab:addlabel1}. 
It is seen from the experimental results of S1  that when a small degree of speech rate mismatch happened, the performance reduced slightly. At the same time, more obvious degradation occurred in the slow test sets than the fast ones, which shows that the slow utterances cause more severe distortion for the spectrum. 
However, when a large degree of speech speed mismatch happened, it could sharply degrade the discriminative power of speaker embeddings, and more degradation occurred in fast speech. One intuitive explanation is that as the audio speed increases, the duration of the audio  becomes shorter and contains less information.

As expected, S2 shows that adding simulated data can alleviate the mismatch problem and improve the performance significantly. 
The S3 and S4 show that feature decomposition  or adversarial learning can further improve performance. The two systems both achieved about 15\% performance improvement in all cases compared to the S2, which trained directly on augmented data. One can also see that in most cases, S4 outperforms S3, while S3 provides better robustness when the speaking rate mismatch became serious.
In S5, we achieved  the best results by combining feature decomposition and cosine similarity adversarial learning, which suggests the two methods are complementary. Surprisingly,  our proposed method not only improves the performance in the mismatch case but also improves the performance in the normal audio speed test set. This observation indicates that the original embeddings contain speech rate information, which would  affect the verification performance, while our proposed method could eliminate speech speed information and improve the discrimination ability for speaker embeddings.

\subsection{Evaluations on HI-MIA}
\label{sub:HI-MIA}

In this section, experiments were conducted on the real scene speaking rate dataset, HI-MIA. 
 We firstly conducted experiments to evaluate whether the TSM based simulated data could characterize the real-world speech rate variability  in the speaker verification task. 
To this end, we utilized the following four different evaluation systems:
\begin{itemize}
  \item S6: The baseline system trained on the HI-MIA slow, normal, and fast speed utterances.
  \item S7: The system  trained only on the HI-MIA normal rate utterances.
  \item S8: The system trained on the  HI-MIA normal rate data combined with two-fold  noise augmented copies. The noise data augmentation processing was followed the setup described in \cite{snyder2018x}.
  \item S9: The system trained on the  HI-MIA normal rate data with two-fold TSM based rate augmentation versions, the $\alpha$  was set to 0.8, 0.9, 1.1, and 1.2.
\end{itemize}
 
 Experimental results of are shown in the upper part of Table~\ref{tab:miya}. 
 We can see from the comparisons between S8 and S9 that S8 outperforms S9 in the normal rate test set, but S9 achieves better preferences under speaking rate mismatch conditions. Since the two systems were trained on the same training data size, it can be deduced that the improved performance achieved by S9 was benefited from the TSM based data augmentation.  Besides, the performance of S9 is nearly similar to S6, which indicates that TSM based data augmentation could characterize the real-world speech rate variability and
 alleviate the real-world speaking rate mismatch problem.

Then, we conducted experiments based on systems of S10-S12 to validate the effectiveness of our proposed method in solving real scene speaking rate mismatch problems. The training datasets of system S10-S12 were the same as the training set of S6. The network structures of S10-S12 correspond to S3-S5.
The lower part of Table~\ref{tab:miya} shows the results by feature decomposition and adversarial learning. We can observe that the experimental results were consistent with those on VoxCeleb1 test sets, which illustrates that our proposed method could also improve the speaker embeddings robustness in real-world speaking rates mismatch scenarios.

\begin{table}[t]
  \centering
  \caption{EER[\%] for verification on HI-MIA.}
  \resizebox{\linewidth}{!}{
    \begin{tabular}{l|ccc|c}
      \hline
      \multirow{2}{*}{Systems} & \multicolumn{3}{c|}{Speaking rate} & \multirow{2}{*}{Average} \\
      & Slow  & Normal  & Fast  &  \\
      
      \hline
      \hline
    
      S6:Baseline  & 1.25  & 1.13  & 4.64  & 2.34  \\
      S7:Normal & 2.51  & 1.89  & 5.26  & 3.22  \\
      S8:Normal + Noise aug. & 2.26  & 1.25 & 5.01  & 2.84  \\
      S9:Normal + TSM aug.   & 1.50  & 1.38  & 4.76  & 2.55  \\
      \hline
      S10:FD att.   & 1.13  & 0.75  & 3.38  & 1.75  \\
      S11:AL cos. & \textbf{1.00} & 0.75  & 2.63  & 1.46  \\
      S12:FD-AL & \textbf{1.00} & \textbf{0.63} & \textbf{2.38} & \textbf{1.34} \\
    \hline

    \end{tabular}}%
  \label{tab:miya}%
\end{table}%

\section{Conclusions}
In this paper, we proposed  an attention  block and cosine similarity loss to obtain rate-invariant speaker embeddings.  
The attention block was used to
obtain rate-related feature and then this feature was disentangled  from the original embedding. The cosine similarity loss and cosine mapping block were introduced to  minimize the  cosine similarity between identity- and rate- related features adversarially.
Experiments conducted on both  VoxCeleb1 based simulated data and  the HI-MIA realistic dataset demonstrated the superior effectiveness of our proposed method in dealing with the speaking rate mismatch problem. In the future, we are interested in validating our method on other speaker-unrelated variabilities, such as far-field, cross-channel, and noisy environments.


\bibliographystyle{IEEEtran}

\bibliography{mybib}

\begin{thebibliography}{10}
\providecommand{\url}[1]{#1}
\csname url@samestyle\endcsname
\providecommand{\newblock}{\relax}
\providecommand{\bibinfo}[2]{#2}
\providecommand{\BIBentrySTDinterwordspacing}{\spaceskip=0pt\relax}
\providecommand{\BIBentryALTinterwordstretchfactor}{4}
\providecommand{\BIBentryALTinterwordspacing}{\spaceskip=\fontdimen2\font plus
\BIBentryALTinterwordstretchfactor\fontdimen3\font minus
  \fontdimen4\font\relax}
\providecommand{\BIBforeignlanguage}[2]{{%
\expandafter\ifx\csname l@#1\endcsname\relax
\typeout{** WARNING: IEEEtran.bst: No hyphenation pattern has been}%
\typeout{** loaded for the language `#1'. Using the pattern for}%
\typeout{** the default language instead.}%
\else
\language=\csname l@#1\endcsname
\fi
#2}}
\providecommand{\BIBdecl}{\relax}
\BIBdecl

\bibitem{snyder2018x}
D.~Snyder, D.~Garcia-Romero, G.~Sell, D.~Povey, and S.~Khudanpur, ``X-vectors:
  {Robust} dnn embeddings for speaker recognition,'' in \emph{Proc.
  ICASSP}.\hskip 1em plus 0.5em minus 0.4em\relax IEEE, 2018, pp. 5329--5333.

\bibitem{zhou2020resnext}
T.~Zhou, Y.~Zhao, and J.~Wu, ``{ResNeXt} and {Res2Net} structure for speaker
  verification,'' \emph{arXiv preprint arXiv:2007.02480}, 2020.

\bibitem{desplanques2020ecapa}
B.~Desplanques, J.~Thienpondt, and K.~Demuynck, ``{ECAPA-TDNN}: {Emphasized}
  channel attention, propagation and aggregation in tdnn based speaker
  verification,'' in \emph{Proc. INTERSPEECH}, 2020, pp. 1--5.

\bibitem{peri2020robust}
R.~Peri, M.~Pal, A.~Jati, K.~Somandepalli, and S.~Narayanan, ``Robust speaker
  recognition using unsupervised adversarial invariance,'' in \emph{Proc.
  ICASSP}.\hskip 1em plus 0.5em minus 0.4em\relax IEEE, 2020, pp. 6614--6618.

\bibitem{tai2020seef}
J.~Tai, X.~Jia, Q.~Huang, W.~Zhang, H.~Du, and S.~Zhang, ``{SEEF-ALDR}: {A}
  speaker embedding enhancement framework via adversarial learning based
  disentangled representation,'' in \emph{Annual Computer Security Applications
  Conference}, 2020, pp. 939--950.

\bibitem{Kwon2020IntraclassVR}
Y.~Kwon, S.-W. Chung, and H.-G. Kang, ``Intra-class variation reduction of
  speaker representation in disentanglement framework,'' in \emph{Proc.
  INTERSPEECH}, 2020.

\bibitem{zeng2015learning}
X.~Zeng, S.~Yin, and D.~Wang, ``Learning speech rate in speech recognition,''
  in \emph{Sixteenth Annual Conference of the International Speech
  Communication Association}, 2015.

\bibitem{grimaldi2009speech}
M.~Grimaldi and F.~Cummins, ``Speech style and speaker recognition: {A} case
  study,'' in \emph{Tenth Annual Conference of the International Speech
  Communication Association}, 2009.

\bibitem{van2007speech}
C.~J. van Heerden and E.~Barnard, ``Speech rate normalization used to improve
  speaker verification,'' in \emph{Proceedings of the Symposium of the Pattern
  Recognition Association of South Africa}.\hskip 1em plus 0.5em minus
  0.4em\relax Citeseer, 2007, pp. 2--7.

\bibitem{rozi2016feature}
A.~Rozi, L.~Li, D.~Wang, and T.~F. Zheng, ``Feature transformation for speaker
  verification under speaking rate mismatch condition,'' in \emph{2016
  Asia-Pacific Signal and Information Processing Association Annual Summit and
  Conference (APSIPA)}.\hskip 1em plus 0.5em minus 0.4em\relax IEEE, 2016, pp.
  1--4.

\bibitem{lee2020nec}
K.~A. Lee, H.~Yamamoto, K.~Okabe, Q.~Wang, L.~Guo, T.~Koshinaka, J.~Zhang, and
  K.~Shinoda, ``{NEC-TT} system for mixed-bandwidth and multi-domain speaker
  recognition,'' \emph{Computer Speech \& Language}, vol.~61, p. 101033, 2020.

\bibitem{ma2018short}
Z.~Ma, H.~Yu, W.~Chen, and J.~Guo, ``Short utterance based speech language
  identification in intelligent vehicles with time-scale modifications and deep
  bottleneck features,'' \emph{IEEE transactions on vehicular technology},
  vol.~68, no.~1, pp. 121--128, 2018.

\bibitem{2006.00848}
T.~Roberts and K.~K. Paliwal, ``A time-scale modification dataset with
  subjective quality labels,'' 2020.

\bibitem{ko2015audio}
T.~Ko, V.~Peddinti, D.~Povey, and S.~Khudanpur, ``Audio augmentation for speech
  recognition,'' in \emph{Sixteenth Annual Conference of the International
  Speech Communication Association}, 2015.

\bibitem{villalba2017tied}
J.~Villalba, N.~Br{\"u}mmer, and N.~Dehak, ``Tied variational autoencoder
  backends for i-vector speaker recognition,'' in \emph{Proc. INTERSPEECH},
  2017, pp. 1004--1008.

\bibitem{kang2019adversarially}
W.~H. Kang and N.~S. Kim, ``Adversarially learned total variability embedding
  for speaker recognition with random digit strings,'' \emph{Sensors}, vol.~19,
  no.~21, p. 4709, 2019.

\bibitem{meng2019adversarial}
Z.~Meng, Y.~Zhao, J.~Li, and Y.~Gong, ``Adversarial speaker verification,'' in
  \emph{Proc. ICASSP}.\hskip 1em plus 0.5em minus 0.4em\relax IEEE, 2019, pp.
  6216--6220.

\bibitem{zhou2019training}
J.~Zhou, T.~Jiang, L.~Li, Q.~Hong, Z.~Wang, and B.~Xia, ``Training multi-task
  adversarial network for extracting noise-robust speaker embedding,'' in
  \emph{Proc. ICASSP}.\hskip 1em plus 0.5em minus 0.4em\relax IEEE, 2019, pp.
  6196--6200.

\bibitem{wang2018unsupervised}
Q.~Wang, W.~Rao, S.~Sun, L.~Xie, E.~S. Chng, and H.~Li, ``Unsupervised domain
  adaptation via domain adversarial training for speaker recognition,'' in
  \emph{Proc. ICASSP}.\hskip 1em plus 0.5em minus 0.4em\relax IEEE, 2018, pp.
  4889--4893.

\bibitem{chen2020channel}
Z.~Chen, S.~Wang, Y.~Qian, and K.~Yu, ``Channel invariant speaker embedding
  learning with joint multi-task and adversarial training,'' in \emph{Proc.
  ICASSP}.\hskip 1em plus 0.5em minus 0.4em\relax IEEE, 2020, pp. 6574--6578.

\bibitem{luu2020channel}
C.~Luu, P.~Bell, and S.~Renals, ``Channel adversarial training for speaker
  verification and diarization,'' in \emph{Proc. ICASSP}.\hskip 1em plus 0.5em
  minus 0.4em\relax IEEE, 2020, pp. 7094--7098.

\bibitem{liu2020text}
K.~Liu and H.~Zhou, ``Text-independent speaker verification with adversarial
  learning on short utterances,'' in \emph{Proc. ICASSP}.\hskip 1em plus 0.5em
  minus 0.4em\relax IEEE, 2020, pp. 6569--6573.

\bibitem{hu2018squeeze}
J.~Hu, L.~Shen, and G.~Sun, ``Squeeze-and-excitation networks,'' in
  \emph{Proceedings of the IEEE conference on computer vision and pattern
  recognition}, 2018, pp. 7132--7141.

\bibitem{zhou2019deep}
J.~Zhou, T.~Jiang, Z.~Li, L.~Li, and Q.~Hong, ``Deep speaker embedding
  extraction with channel-wise feature responses and additive supervision
  softmax loss function.'' in \emph{Proc. INTERSPEECH}, 2019, pp. 2883--2887.

\bibitem{wang2019decorrelated}
H.~Wang, D.~Gong, Z.~Li, and W.~Liu, ``Decorrelated adversarial learning for
  age-invariant face recognition,'' in \emph{Proceedings of the IEEE/CVF
  Conference on Computer Vision and Pattern Recognition}, 2019, pp. 3527--3536.

\bibitem{nagrani2020voxceleb}
A.~Nagrani, J.~S. Chung, W.~Xie, and A.~Zisserman, ``Voxceleb: {Large}-scale
  speaker verification in the wild,'' \emph{Computer Speech \& Language},
  vol.~60, p. 101027, 2020.

\bibitem{qin2020hi}
X.~Qin, H.~Bu, and M.~Li, ``{HI-MIA}: {A} far-field text-dependent speaker
  verification database and the baselines,'' in \emph{Proc. ICASSP}.\hskip 1em
  plus 0.5em minus 0.4em\relax IEEE, 2020, pp. 7609--7613.

\bibitem{tomar2006converting}
S.~Tomar, ``Converting video formats with ffmpeg,'' \emph{Linux Journal}, vol.
  2006, no. 146, p.~10, 2006.

\bibitem{snyder2019speaker}
D.~Snyder, D.~Garcia-Romero, G.~Sell, A.~McCree, D.~Povey, and S.~Khudanpur,
  ``Speaker recognition for multi-speaker conversations using x-vectors,'' in
  \emph{Proc. ICASSP}.\hskip 1em plus 0.5em minus 0.4em\relax IEEE, 2019, pp.
  5796--5800.

\bibitem{tong2021asv}
F.~Tong, M.~Zhao, J.~Zhou, H.~Lu, Z.~Li, L.~Li, and Q.~Hong, ``{ASV-Subtools}:
  {Open} source toolkit for automatic speaker verification,'' in \emph{Proc.
  ICASSP}.\hskip 1em plus 0.5em minus 0.4em\relax IEEE, 2021, pp. 6184--6188.

\end{thebibliography}

\end{document}